| **Title** | **Direct current parallel microdischarges in helium** |
|---|---|
| **Authors** | T. Dufour[1], R. Dussart[1], L. J. Overzet[2], P. Lefaucheux[1], P. Ranson[1], J.-B. Lee[2], M. Mandra[2], M. J. Goeckner[2], N. Sadeghi[3] |
| **Affiliations** | [1]GREMI/Université d'Orléans, Orléans, France<br>[2]University of Texas at Dallas, Richardson, TX, USA<br>[3]LSP/Université Joseph Fourier Grenoble & CNRS, Saint-Martin d'Hères, France |
| **Ref.** | ISPC-18, Kyoto, Japan, 26th-31st August 2007<br>https://plas.ep2.rub.de/ispcdocs/ispc18/ispc18/content/paper00524.pdf |
| **DOI** | - |
| **Abstract** | Parallel Micro Hollow Cathodes Discharges are characterized by electrical and spectroscopic studies and by using an ICCD camera. V-I curves present a hysteresis cycle between Townsend and normal glow regimes. For Id=4-10mA, gas temperature ranges between 300-500°C. When the diameter of a hole decreases, a more significant discharge voltage is needed. |

# 1. Introduction

Microdischarges are non equilibrium discharges, spatially confined to dimensions smaller than to 1mm. They are a promising approach to the generation and maintenance of stable glow discharges at atmospheric pressure and represent new challenges for plasma science (possible breakdown of p.d scaling, impact of quantum electrodynamics on spontaneous emission rate, breakdown of quasi-neutrality) [1]. Parallel microplasmas can be created using several kinds of microdevices [2]. Some operate at rather high discharge current (up to 65 mA) without the need of ballasts [3]. Some other teams use individual ballasts to initiate the parallel microplasmas [4]. Our aim is to investigate both single and parallel multi MHCD devices by electrical, spectroscopic and imaging characterisations.

# 2. Experimental setup

The MHCD used in this experiment consists of a Ni:$Al_2O_3$:Ni sandwich structure. A nickel layer of 5 µm thickness, is deposited on both surfaces of an $Al_2O_3$ layer (250µm thickness), by electrodeposition at UTDallas. 135 to 300 µm diameter holes are drilled by a Nd:Yag laser through this structure. We study single MHCD devices but also parallel multi MHCD devices with 7, 12, 20 or 31 holes, separated with various interhole distances from 175 to 380 µm.

The MHCD was inside a vacuum chamber linked to a primary pump, so that a pure gas environment could be maintained and the discharge could be operated below atmospheric pressure, without any gas flow. The experimental set-up is shown in Fig. 1. Experiments are carried out in He, Ar and in He/$N_2$ mixtures, at pressures between 50 and 1000 Torr, and measured with an MKS Baratron gauge 622A (reading accuracy of 0.25%). A 0-2500 V DC bias power supply is used to operate the microdischarges. Electrical measurements are performed using a Tektronix oscilloscope and high voltage probes. The discharge voltage is measured between inlet and output resistances ($R_1$=19kW, 16W and $R_2$=1kW) and the discharge current through R2 connected to ground. Optical emission spectroscopy is carried out with a Jobin Yvon TRIAX 550 spectrometer with 3 different gratings: 250, 1200 and 2400 lines.mm-1. An intensified CCD camera is installed to observe the light emission of the individual MHCD. This is a PI-MAX 512 from Princeton Instruments/Acton with a 512×512 imaging array. We equipped it with a macro zoom lens Nikon Micro-Nikkor 60 mm f/2.8D. Studies are carried out in visible range, since optics and windows are in BK7. For experiments, the ICCD camera is used with a gate width of 50ns, a gate delay of 2ns, 1 gate per exposure and a gain of 250.

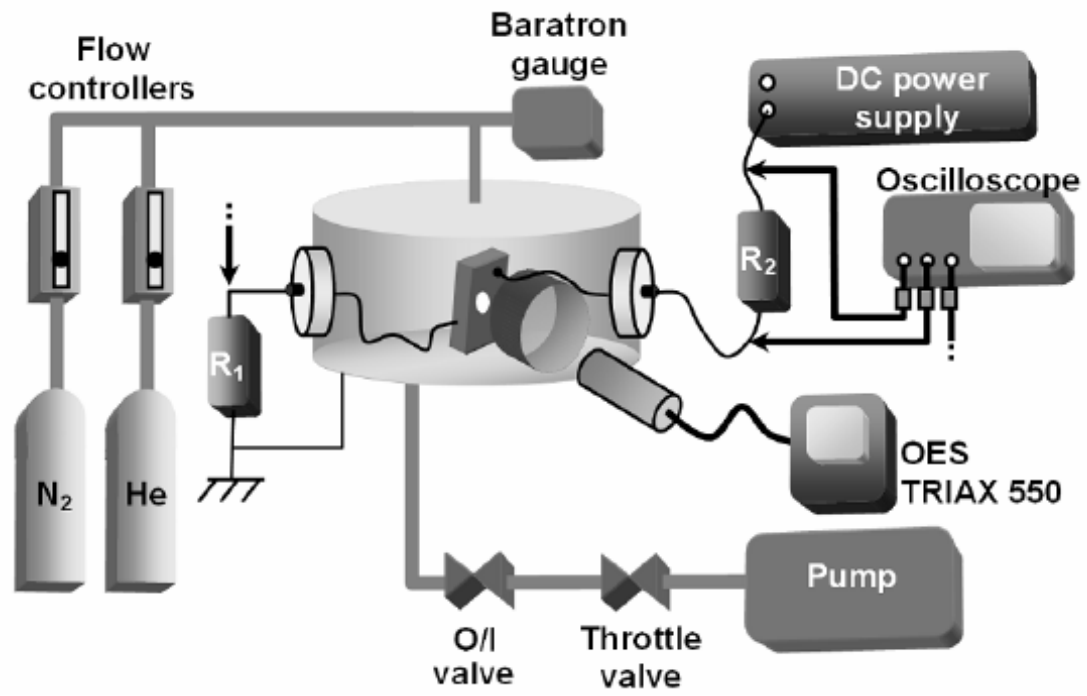

***Figure 1. Experimental setup***

# 3. Results and discussions

## 3.1. Single hole microdevices

#### 3.1.1. V-I curve

A V-I characteristic is shown in Fig. 2 for a single MHCD microdevice (diameter: 300 µm) in He for a pressure of 400 Torr. Three different regions are clearly identified: the pre-breakdown, the normal glow and the self-pulsing regime. This self-pulsing regime was observed and studied in detail by another group [5].

#### 3.1.2. Influence of the hole diameter on discharge voltage

By applying a voltage ramp to the microdischarge, one can measure the V-I characteristics of the microdevice. The V-I curves for 130µm and 300µm diameter MHCDs are shown in Fig. 3 for 600 Torr He. One can see on this graph that the breakdown voltages (315 and 300V) are much larger than the operating voltages (approximately 180 and 170V). The big difference between breakdown and operating voltages implies that in the case of multiple hole device, once a first hole has ignited, the voltage will become too small for a second hole to breakdown independently. This implies it will be hard to light multiple holes at once with the present simply fabricated devices.

The V-I characteristic between 2 and 15 mA has a slightly negative slope, corresponding to the typical normal glow discharge. In a normal cylindrical glow, the solution of the diffusion equation [6] results in a characteristic diffusion length (L):

$$\frac{1}{\Lambda^2} = \left(\frac{2.405}{D/2-s}\right)^2 + \left(\frac{\pi}{\ell-2s}\right)^2 = \frac{\nu_{iz}}{D_a}$$

This equation relates the ratio of the ionization rate coefficient ($\nu_{iz}$) to the ambipolar diffusion coefficient ($D_a$). We assume $D_a$ varies slowly compared to $\nu_{iz}$. We note that s represents the sheath width correction. According to Eq. (1), increasing hole diameter (D) corresponds to a decrease of $\nu_{iz}$ and therefore of the electric field in the positive column of the normal glow. This explains why a larger D requires a smaller $\nu_{iz}$ and electric field. As a result, a smaller discharge voltage is needed, as observed in Fig. 3.

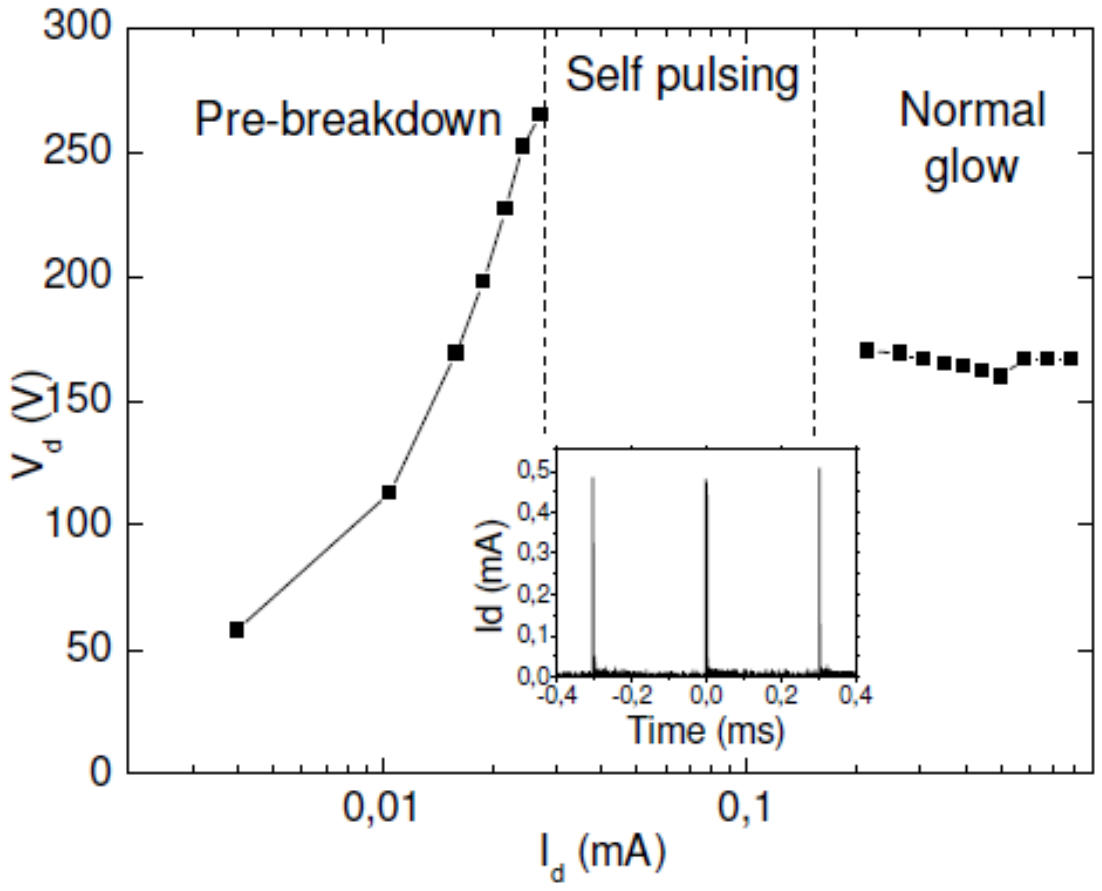

***Figure 2. IV curve in He (400 Torr)***

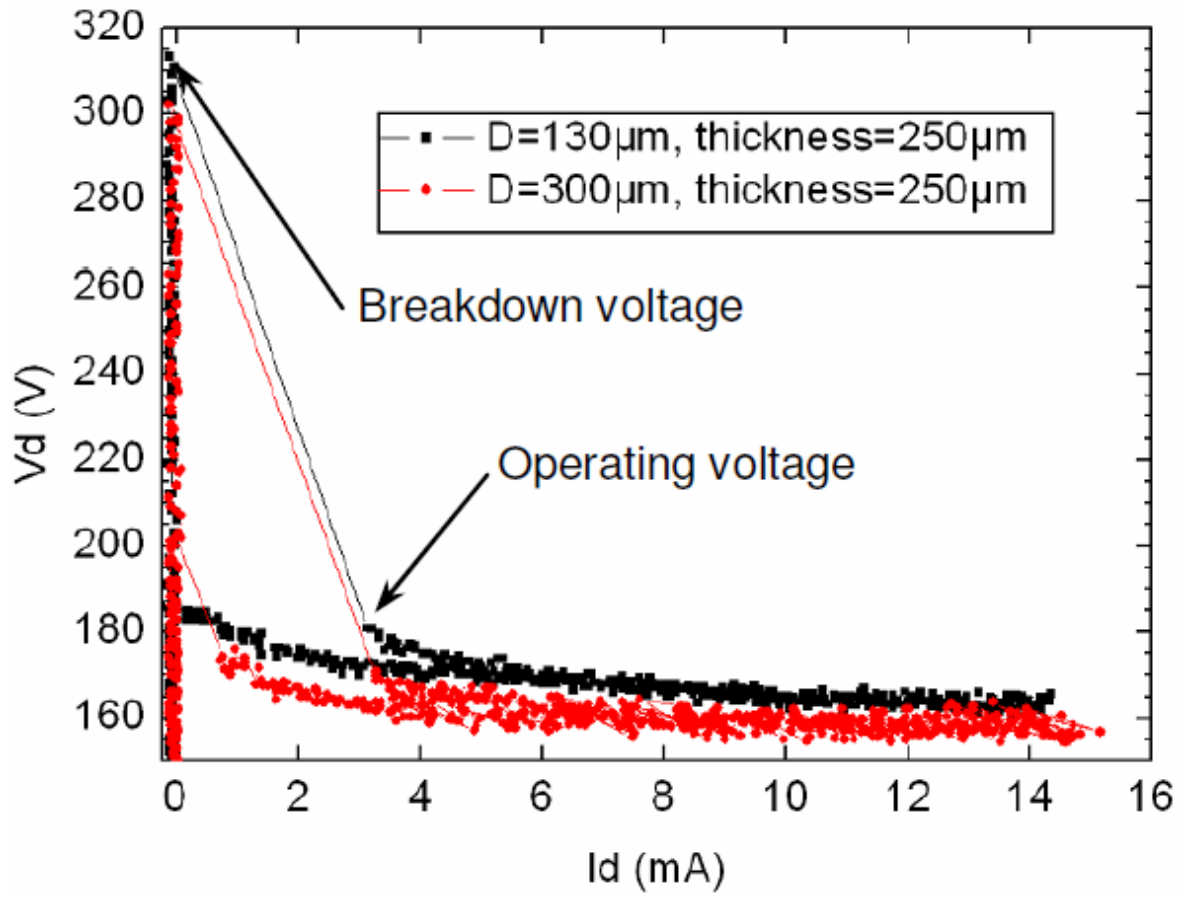

***Figure 3. V-I curves for two diameters of single MHCD microdevices.***

**3.1.3. Influence of the hole diameter on light emission**

We measured the light emitted from different hole diameters using the ICCD camera. We obtain a picture representing the light emitted from each hole. The each picture is analyzed using a Matlab program, which designs a circular test zone and only takes in to consideration "active" pixels. Active pixels are defined as those having light intensity superior to a minimum level set by noise. These active pixels are represented on Fig. 4 by green points. The integrated light intensity of a hole is the sum of all active pixel intensities. The integrated light intensity is plotted as a function of the discharge current for 3 different MHCD diameters in Fig. 5. The light intensity monotonically increases as a function of the discharge current and for increasing hole diameter.

According to Fig. 3, the voltage across the MHCD is nearly independent of the current. As a consequence, the power is linearly related to the current. Therefore the integrated light intensity simply increases with the power deposited inside the glow. If one divides the integrated light intensity by the hole area, one gets an average intensity that is the same order of magnitude for all three hole sizes. This result implies that the light intensity should remain quite significant for even smaller holes.

**3.1.4. Determination of gas temperature**

By adding a small amount of $N_2$ to the He gas, it is possible to deduce the gas temperature by fitting the second positive system $C^3\Pi_u$-$B^3\Pi_g$ emission spectra with the intensity of the general rotational, vibronic transition Cv'J'-Bv"J" (Fig. 6):

$$F_{Bv"J"}^{Cv'J'} = M\frac{S_j}{T}\exp\left[-\frac{hcB_v}{kT}J'(J'+1)\right]$$

where T: gas temperature, h : Planck constant, k : Boltzman constant, M: coefficient defined by Porter and Harshbarger [6], and $S_j$ the general line strength taken from the book of Kovacs [7]. The fitting precision is about ±20°C.

For several discharge currents, Fig. 7 shows the increase of the gas temperature, with Id up to 500°C for a 10 mA discharge current. The gas temperature is well below an arc temperature, which indicates that the MHCD is in normal glow regime and agrees with V-I characteristic. The gas temperature increases linearly with the discharge current and power. This is to be expected since the power is deposited through the DC electric field. The gas temperature inside the MHCD is much larger than the ambient. As a consequence, a significant gradient in atom density is expected to develop as the plasma ignites. A flow should be obtained between the outer region and the cavity.

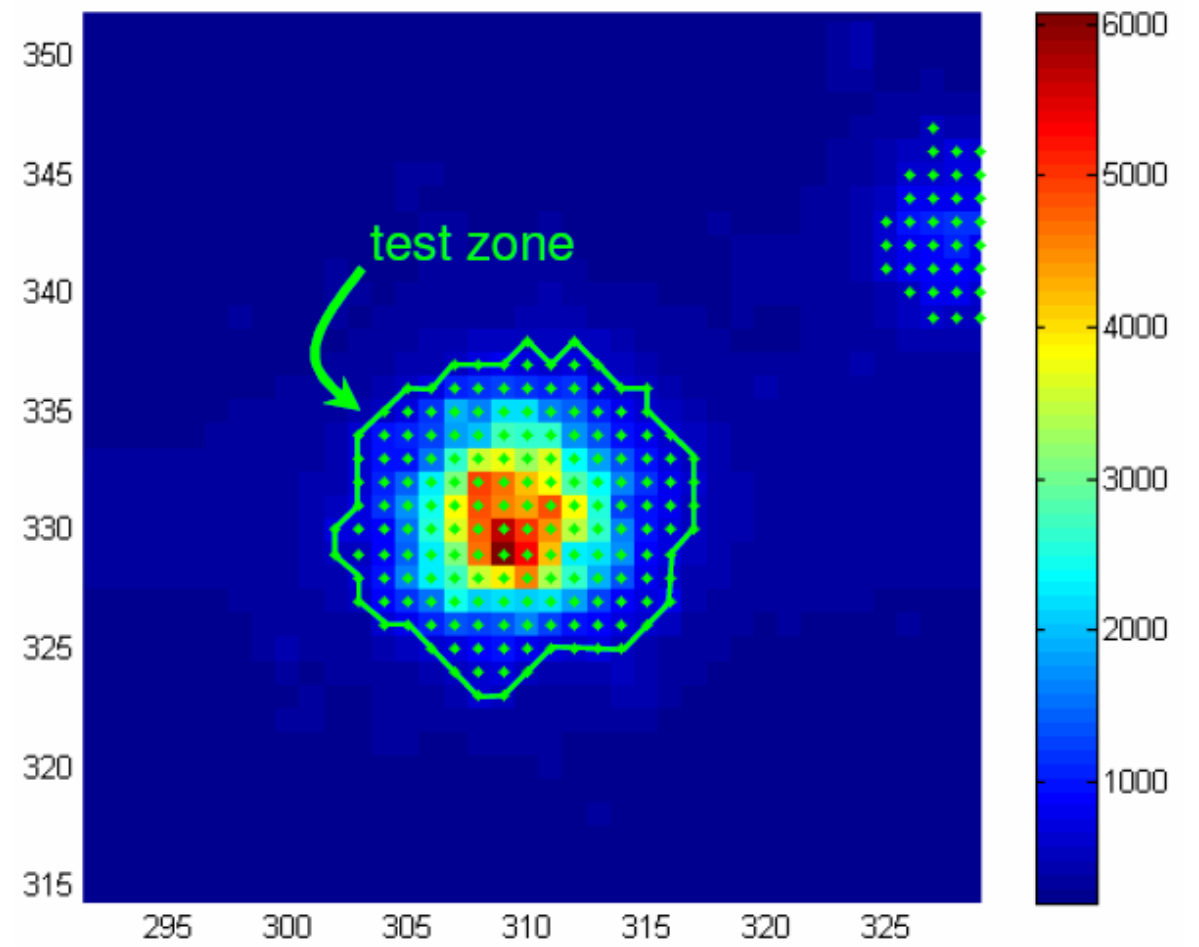

***Figure 4. Picture of a microdischarge. Green points denote the pixels, which are summed.***

***Figure 5 Average light intensity emitted by single microdischarges, for different hole diameters.***

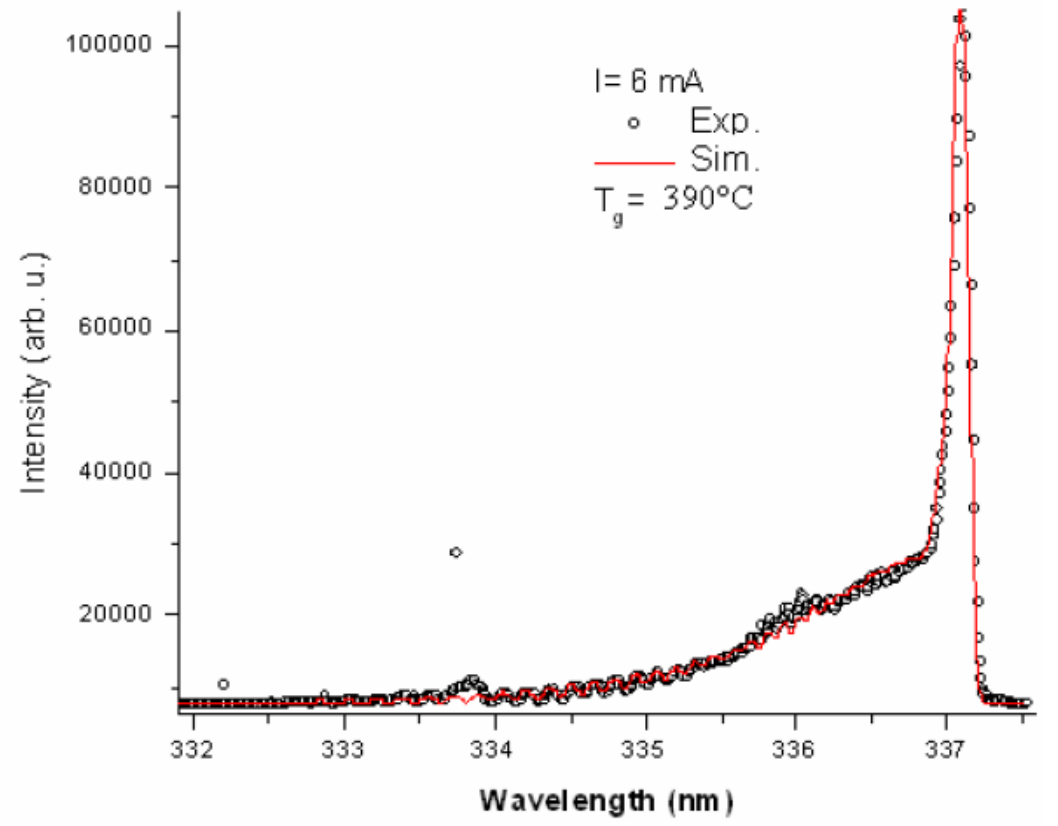

***Fig. 6 Fitting of $N_2$ 2nd positive system $C^3\Pi_u$-$B^3\Pi_g$ (OES) for $I_d$=6mA.***

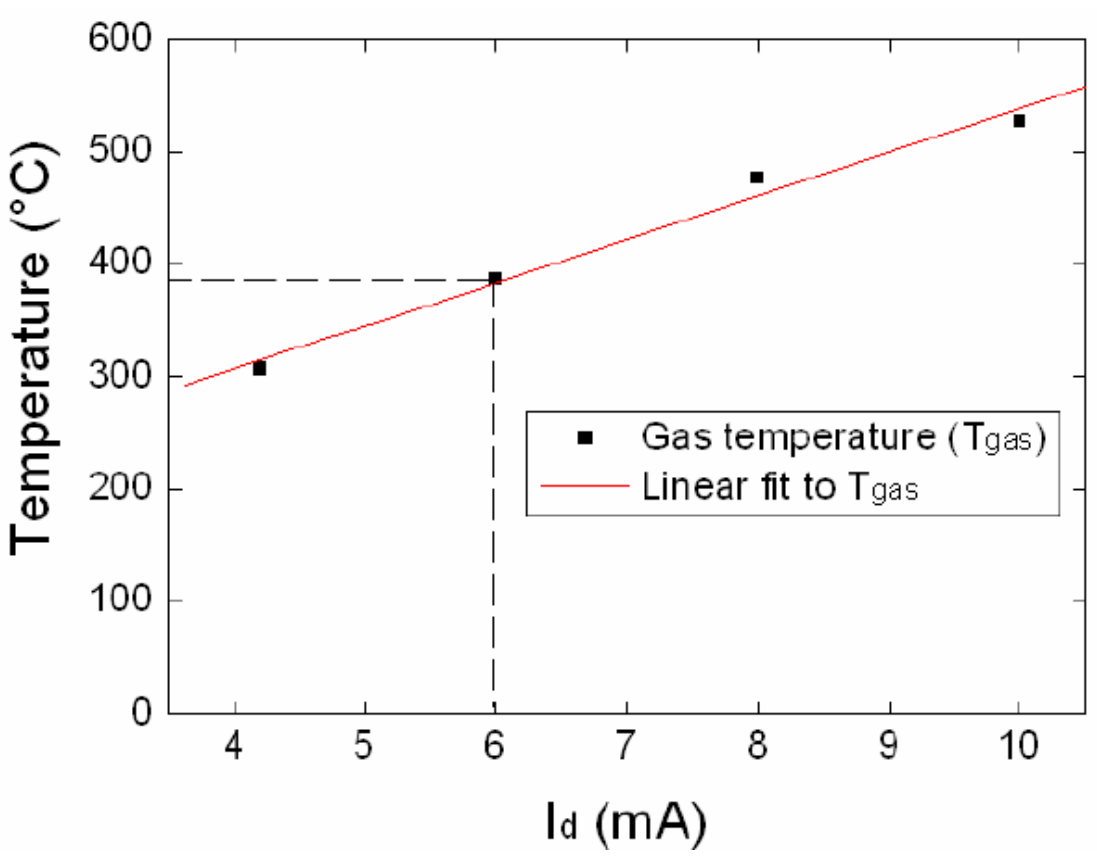

***Figure 7. Gas temperature for $P_{N2}$=7 Torr and $P_{He}$=440 Torr.***

## 3.2. Multi-MHCD microdevices

An experiment was carried out in He (400 Torr), for a 12 MHCD microdevice with hole diameters of 200µm and hole spacings of 300µm. As presented in the V-I curve (Fig. 8), the discharge voltage ranges between 160V and 190V. For each datapoint, a picture of the MHCD was taken from the anode side. The inset is the same V-I curve, but shows the near constant voltage as a function of current. In the upper left corner is a picture of the micro device without plasma for reference.

For $I_d$ between 2 and 30mA, $V_d$ decreases whereas the number of working microdischarges increases. On the contrary, for Id=30-40 mA, $V_d$ increases while the number of working microdischarges decreases back to one. Our parallel microdischarges always began by the ignition of a single hole. The ignition of a first hole corresponds to a decrease of the voltage across the microdischarge. This makes the voltage across the other cavities sit well below the breakdown voltage.

As the discharge current is increased from low values, the number of operating microdischarges also increased from one to as many as five. In the light of the V-I characteristic of the one hole microdischarge, this result is surprising. The negative differential resistance indicates that only one hole should carry all the current. In addition, we find that only the holes of the upper left part of the micro device operate for this device. Normal glow microdischarge spreads along the outer cathode surface while the current is increased [5]. When this microdischarge extension reaches one of the other holes with sufficient density, another hole can be initiated, as observed in Fig. 8.

The second hole initiated does not have to be the closest to the first. Some reasons can be advanced such as surface oxidation, roughness and scratches on the cathode surface. As a consequence, a close hole spacing is a necessary but not sufficient condition for parallel microdischarges working in the normal glow regime. The reasons why the number of lit microplasmas decreases at high current is not yet understood. It may be that the gas temperature in the holes become different enough so that the E/N value becomes different enough and the ionization rate in one hole dominates.

In another experiment (600 Torr He in a device having 20 cavities of 130 µm diameter), we also followed the integrated light intensity emitted by each hole as a function of the discharge current (Fig. 9). For $I_d$=5mA, only one microdischarge is on. We call it "hole N°01 ()". At $I_d$=13mA, the light intensity of N°01 suddenly decreased, whereas N°03 () turns on abruptly and N°04 () increases progressively with current. Above 20 mA there are at least 3 cavities (N°02, 03, 04) which have about the same integrated intensity. At 38 mA, all of the plasma has coalesced into just one hole again (N°05). Multiple microdischarge devices operating in normal discharge mode can have similar emission intensities. The light intensity from a specific hole does not necessarily increase with $I_d$.

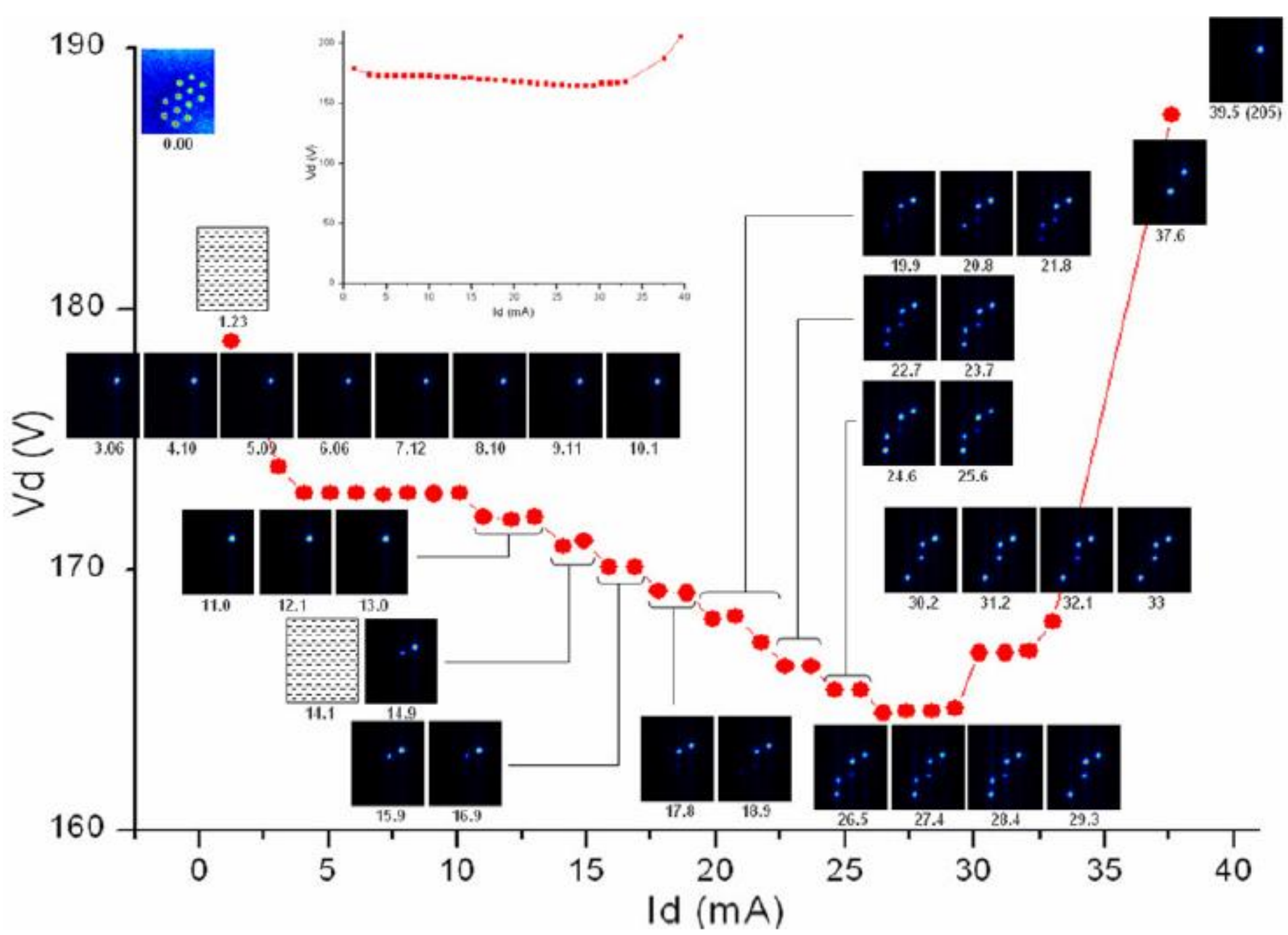

***Fig. 8 Typical V-I curve for a 12 MHCD microdevice along with corresponding pictures of the light emission The V-I curve inset (upper left) shows the full voltage range. The picture in the upper left corner shows the locations of the 12 holes.***

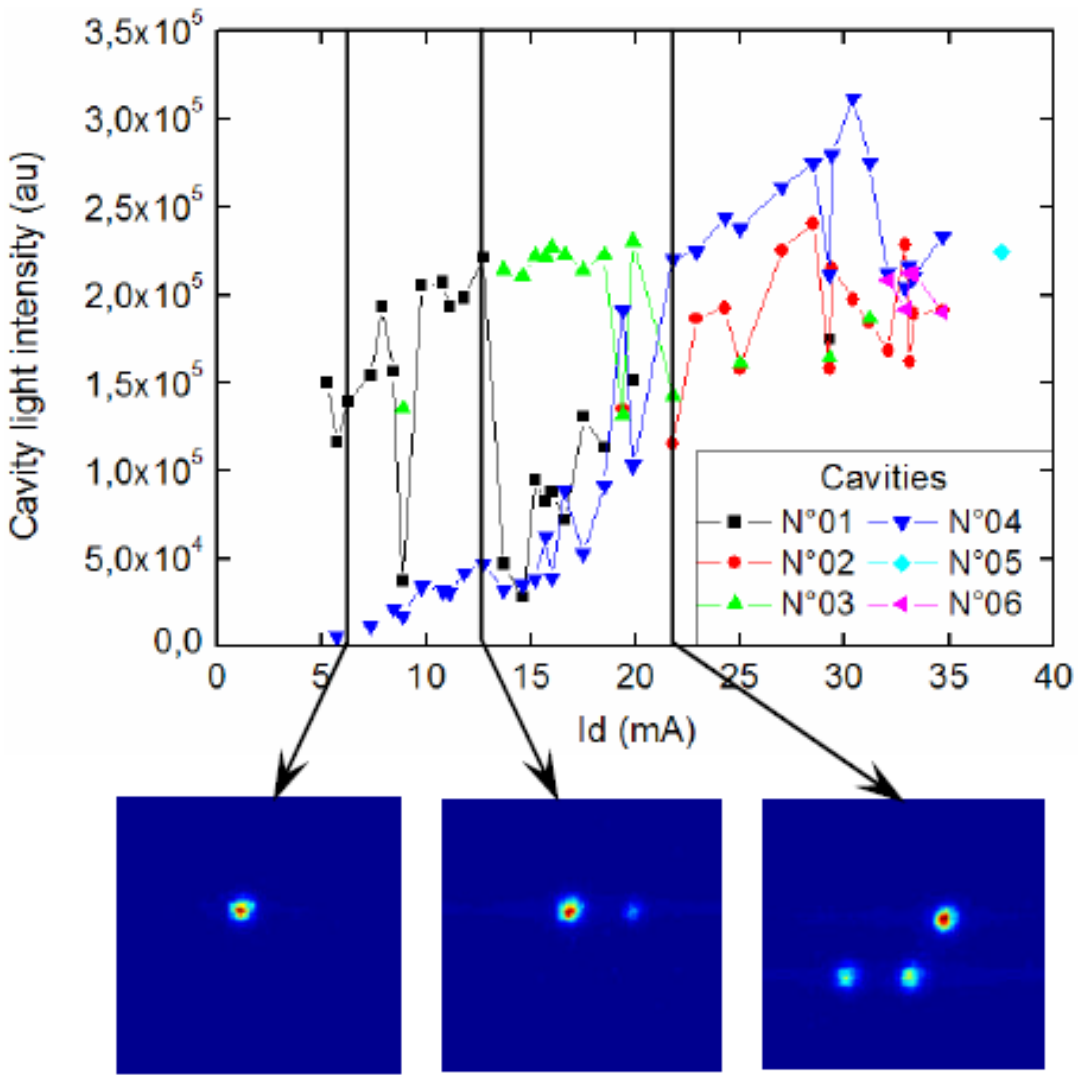

***Figure 9. Order of microdischarges ignitions versus Id. Observation on the anode surface.***